\documentclass[letterpaper,10pt,twocolumn]{article}

\usepackage[latin1]{inputenc}
\usepackage{amsmath}

\usepackage[margin=15mm]{geometry}
\usepackage{titlesec}
\usepackage[labelfont=bf,justification=raggedright]{caption}

\usepackage{graphicx}
\usepackage{epstopdf}
\usepackage{color}

\usepackage[sort&compress,comma]{natbib}
\usepackage{hyperref}

\titleformat{\section}{\bfseries}{\thesection}{0em}{}	
\titleformat{\subsection}[runin]{\normalfont\bfseries}{}{0em}{}	
\titlespacing*{\section}{0pt}{0pt}{-\parskip}

\newcommand{\bcaption}[2]{\caption{\textbf{#1} #2}}

\setcitestyle{super,open={},close={}}\bibfont{\small}
\renewcommand{\bibfont}{\normalfont\small}
\hypersetup{colorlinks=true,allcolors={blue},citecolor={red}}

\DeclareGraphicsExtensions{.eps,.pdf}

\begin{document}


\begin{titlepage}

\vspace*{3cm}
\begin{center}
\LARGE{\textbf{Nonlinear optics of fibre event horizons}}
\vspace*{1cm}

\large{Karen~E.~Webb$^\text{1}$, Miro~Erkintalo$^\text{1*}$, Yiqing~Xu$^\text{1,2}$, Neil~G.~R.~Broderick$^\text{1}$, John~M.~Dudley$^\text{3}$, Go\"ery~Genty$^\text{4}$, \& Stuart~G.~Murdoch$^\text{1}$}
\vspace*{5mm}

\textit{\small{
$^\text{1}$Department of Physics, University of Auckland, Private Bag 92019, Auckland, New Zealand.\\
$^\text{2}$Department of Electrical and Electronic Engineering, University of Hong Kong, Pokfulam Road, Hong Kong.\\
$^\text{3}$Universit\'e de Franche-Comt\'e, Institut FEMTO-ST, CNRS UMR 6174, 25030 Besan\c{c}on, France.\\
$^\text{4}$Optics Laboratory, Department of Physics, Tampere University of Technology, P.O.B 692, Tampere, Finland.\\
Corresponding author: m.erkintalo@auckland.ac.nz}}
\end{center}
\vspace*{1cm}

The nonlinear interaction of light in an optical fibre can mimic the physics at an event horizon. This analogue arises when a weak probe wave is unable to pass through an intense soliton, despite propagating at a different velocity. To date, these dynamics have been described in the time domain in terms of a soliton-induced refractive index barrier that modifies the velocity of the probe. Here, we complete the physical description of fibre-optic event horizons by presenting a full frequency-domain description in terms of cascaded four-wave mixing between discrete single-frequency fields, and experimentally demonstrate signature frequency shifts using continuous wave lasers. Our description is confirmed by the remarkable agreement with experiments performed in the continuum limit, reached using ultrafast lasers. We anticipate that clarifying the description of fibre event horizons will significantly impact on the description of horizon dynamics and soliton interactions in photonics and other systems.

\end{titlepage}
\setlength{\parindent}{15pt}
\setlength{\parskip}{0pt}
\noindent The demonstration that an intense pulse propagating in an optical fibre can give rise to an artificial event horizon \cite{philbin08} has attracted considerable attention with the possibility to produce laboratory white- or black-hole analogues \cite{philbin08,hill09,faccio10,robertson10,demircan11,choudhary12, demircan12,faccio12a,faccio12b,tartara12}. This picture not only suggests interesting links with general relativity \cite{philbin08,faccio10,robertson10,faccio12a,faccio12b}, but the underlying dynamics have also been postulated to influence the formation of optical rogue waves \cite{solli07,demircan12,demircan13}, and to allow for the realization of optical functionalities such as the all-optical transistor \cite{demircan11}.

A fibre-optical analogue of an event horizon occurs when a bright soliton prevents a low power probe, travelling at a different velocity, from passing through the soliton. The underlying physics has been described in terms of the intensity dependence of the fibre refractive index (Kerr effect): the soliton creates a moving refractive index perturbation that alters the velocity of the probe wave, preventing its passage through the soliton\cite{philbin08, faccio12a, belgiorno10}. To the probe, the soliton boundary can thus appear as a horizon which light can neither enter nor escape, in analogy with behaviour at the boundary of a white- or black-hole\cite{philbin08}. In a dispersive medium, such as an optical fibre, a change in velocity must be associated with a change in frequency, which has been attributed to the conservation of the Doppler shifted probe frequency \cite{philbin08, faccio12a}. This allows signatures of horizon dynamics to be identified in the probe spectrum, where frequency up-conversion (blue-shift) occurs at a white-hole horizon, and frequency down-conversion (red-shift) occurs at a black-hole horizon\cite{philbin08,robertson10,choudhary12,faccio12b}. Of course, the time- and frequency-domain descriptions represent identical physics; a change in probe velocity can be equally understood to arise from the generation of new frequencies, linking the dynamics of fibre event horizons to the realm of nonlinear optical frequency conversion \cite{agrawal_nlfo, skryabin10}.

So far, all reported studies have focused on ultrashort pulses when considering fibre event horizons. However, the spectrum of a train of such pulses is a frequency comb that consists of discrete spectral components \cite{eckstein78,udem02}, suggesting that the dynamics must possess an equivalent description that relies solely on the nonlinear mixing of monochromatic continuous waves (CWs) \cite{erkintalo12,boyd}. In this work, we unveil this description by interpreting the dynamics of fibre-optical event horizons in terms of interactions between a discrete set of monochromatic continuous waves, and by experimentally demonstrating horizon signatures using only CW lasers. First, we summarize the direct correspondence between fibre horizons and established supercontinuum frequency conversion mechanisms \cite{skryabin10}, then explain the underlying frequency conversion dynamics in terms of cascaded four-wave mixing~(FWM) triggered by a set of CW fields. Our analysis is confirmed by experiments and simulations seeded by both pulsed and CW lasers, over the full range of observable event horizon dynamics. Since a discrete set of modes constitutes the basis of any quantum field theory, our results could open entirely new ways of analysing analogue event horizons and soliton interactions. Significantly, our FWM-based description intrinsically allows for the underlying energy conservation and photon-pair generation mechanisms to be identified, paving the way for the conclusive observation of the photonic analogue of Hawking radiation \cite{belgiorno10,barcelo11, petev13, unruh14}.

\begin{figure*}[t!]
\centering
\includegraphics[clip=true]{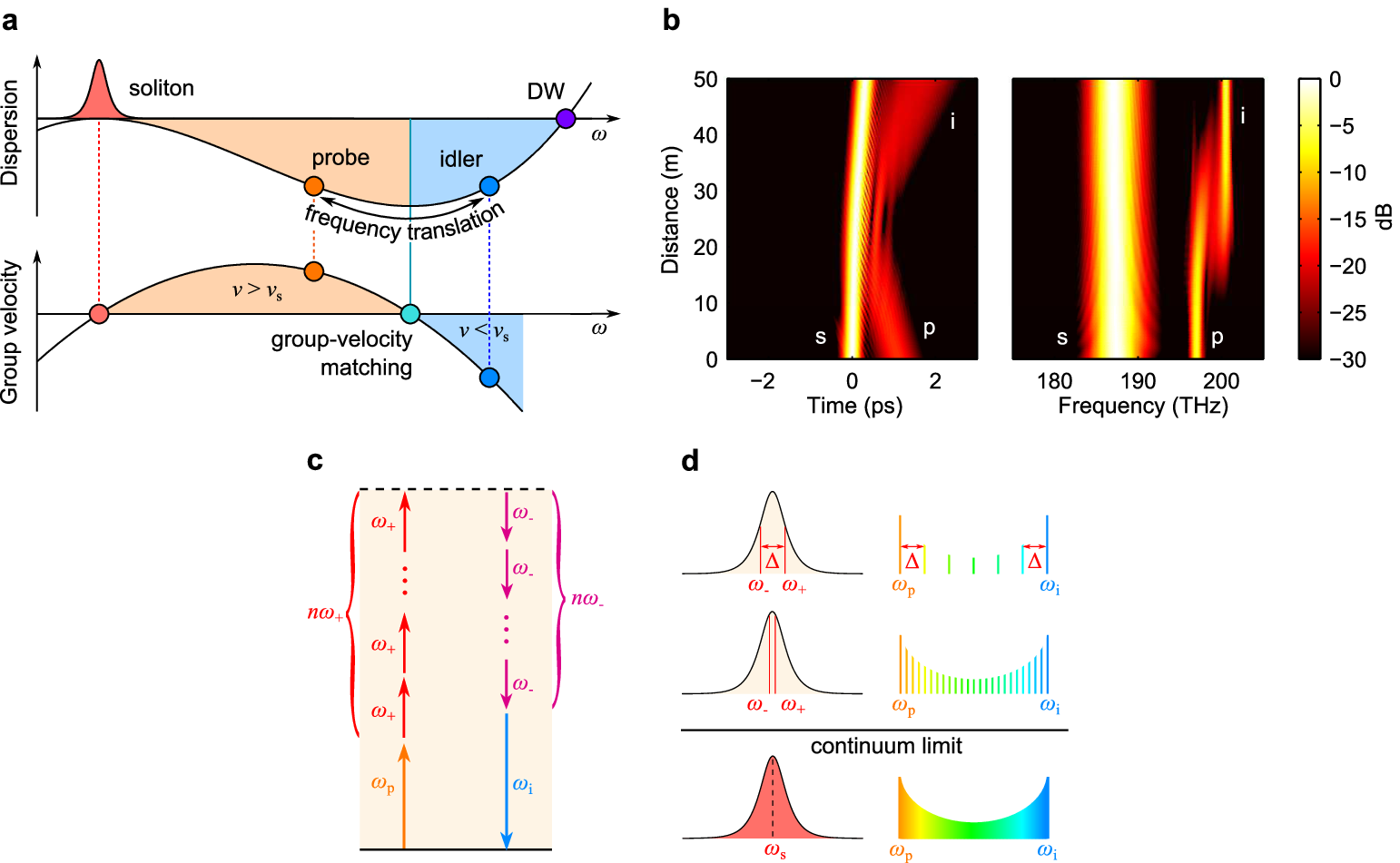}
\bcaption{Illustration of the nonlinear optics of fibre horizons.}{(\textbf{a})~Dispersion (top) and corresponding group velocity (bottom) illustrate how phase-matched probe-idler pairs experience group velocities with opposite signs in the soliton reference frame. In the laboratory reference frame, waves whose frequencies lie in the orange (blue) shaded area propagate faster (slower) than the soliton. (\textbf{b})~Density maps of the temporal (left) and corresponding spectral (right) evolution of a soliton pump and probe pulse in an optical fibre (s$=$soliton, p$=$probe, i$=$idler). In the time domain, the probe appears to reflect off the soliton, owing to the generation of an idler at $\omega_\text{i}$. (\textbf{c})~Energy-level description of the nonlinear process mimicked by the four-wave mixing cascade. For a probe frequency above the group-velocity matched point, the frequencies~$\omega_+$ and~$\omega_-$ are swapped. (\textbf{d})~All monochromatic frequency pairs symmetrically detuned around the soliton centre induce a discrete comb that transfers energy to the same higher-order idler. In the continuum limit, the soliton picture is recovered.}
\label{schematic}
\vspace{20pt}
\end{figure*}


\section*{Results}
\subsection{Event horizons in nonlinear fibre optics.} The generation of new frequencies through interactions of intense solitons and weak linear waves was first predicted theoretically by Yulin et al. \cite{yulin04,skryabin05}, and experimentally observed in the context of fibre supercontinuum generation by Efimov et al. \cite{efimov04,efimov05,chapman10}, although the link to analogue event horizons was not identified at the time. Recent theoretical work\cite{choudhary12,demircan12,faccio12a,driben13,yulin13} has alluded to the connection between these nonlinear interactions and fibre event horizons, but the relationship is yet to be explicitly demonstrated. Formally, however, the resonant interaction of solitons and weak linear waves describes the very same physics that underpins fibre-optic analogues of event horizons. As this result is important for our CW analysis, we begin by clarifying the equivalence of the two phenomena.

Consider a soliton centred at an optical frequency~$\omega_\text{s}$, propagating in the anomalous dispersion regime, together with a weak probe at frequency~$\omega_\text{p}$ in the normal dispersion regime. Due to their nonlinear interaction, the probe can experience a frequency conversion to an idler $\omega_\text{p}\rightarrow\omega_\text{i}$. The idler frequency~$\omega_\text{i}$ is set by the resonance condition \cite{yulin04,skryabin05, skryabin10, chapman10}
\begin{equation}
\hat{D}(\omega_\text{i}-\omega_\text{s})=\hat{D}(\omega_\text{p}-\omega_\text{s}),
\label{resonance1}
\end{equation}
where $\hat{D}=\sum_{k\ge2}{\beta_k(\omega-\omega_\text{s})^k/k!}$ is the dispersion operator in the reference frame of the soliton, representing the frequency-dependence of the fibre refractive index ($\beta_2$ is the group-velocity dispersion coefficient and \mbox{$\beta_k$ ($k=3,4,\dots$)} are higher-order dispersion coefficients). This operator is obtained from the Taylor series expansion of the dispersion around~$\omega_\text{s}$ (removing the zeroth and first order terms, $\beta_0=\beta(\omega_\text{s})$ and $\beta_1=d\beta/d\omega|_{\omega_\text{s}}$, respectively) \cite{agrawal_nlfo}. We first note that in the degenerate case where~$\omega_\text{p}=\omega_\text{s}$, Eq.~\eqref{resonance1} describes the emission of a Cherenkov dispersive wave (DW) by a soliton \cite{wai86,akhmediev95, skryabin10}. In the non-degenerate case, where~$\omega_\text{p}\neq\omega_\text{s}$, the nonlinear frequency conversion $\omega_\text{p}\rightarrow\omega_\text{i}$, governed by Eq.~\eqref{resonance1}, manifests itself as an event horizon in the time domain. This is illustrated in Fig.~\ref{schematic}a-b. In Fig.~\ref{schematic}a, the top graph shows the cubic dispersion profile, $\hat{D}(\omega-\omega_\text{s})$, of a typical telecommunications fibre, while the bottom graph shows the corresponding group velocity, $v_\text{g}=1/\beta_1(\omega-\omega_\text{s})$. In Fig.~\ref{schematic}b, we show an illustrative simulation of the temporal (left panel) and spectral (right panel) evolution of a soliton pump and a weak probe pulse propagating in the fibre (see Methods). Note that, as is customary, this simulation is performed in a reference frame moving with the soliton, which therefore appears stationary in the time domain. We can see from Fig.~\ref{schematic}a that any probe-idler pair, located between the DW and pump frequencies and connected through the resonance condition in Eq.~\eqref{resonance1}, will experience group velocities of opposite sign relative to the soliton. This means that, when interacting with the soliton, a probe travelling faster than the soliton ($v_\text{p}>v_\text{s}$) will generate an idler travelling slower than the soliton ($v_\text{i}<v_\text{s}$), and vice versa.  In the time domain reference frame of the soliton, the frequency conversion therefore gives rise to an apparent ``reflection'' of the probe off the soliton\cite{demircan11,rubino12b,yulin13}, highlighted in Fig.~\ref{schematic}b. The probe, travelling faster than the soliton, is seen to bounce off from the soliton at a distance of~$\sim25$~m (left panel); or equivalently to generate a frequency-shifted idler that travels slower than the soliton. Indeed, the right panel shows the corresponding spectral evolution of the field, and clearly demonstrates that the interaction is associated with the frequency conversion~$\omega_\text{p}\rightarrow\omega_\text{i}$. Interchanging the roles of the probe and the idler simply change the nature of the event horizon: the frequency that is group-velocity matched to the soliton represents the transition point between white- and black-hole horizons. The impenetrability of the soliton (event horizon) and the nonlinear fibre-optic frequency conversion therefore describe the very same physics; the former describes a change in probe velocity whose signature is a frequency translation, while the latter describes a frequency translation whose signature is a change in velocity. Finally, we remark that a fundamental property of the frequency conversion mechanism described above is that it manifests itself only when the fibre group-velocity dispersion varies with frequency (higher-order dispersion coefficients are nonzero)\cite{yulin04,skryabin05, skryabin10}, which naturally now implies the same requirement for the observation of event horizon dynamics.

\begin{figure*}[t!]
\centering
\includegraphics[clip=true]{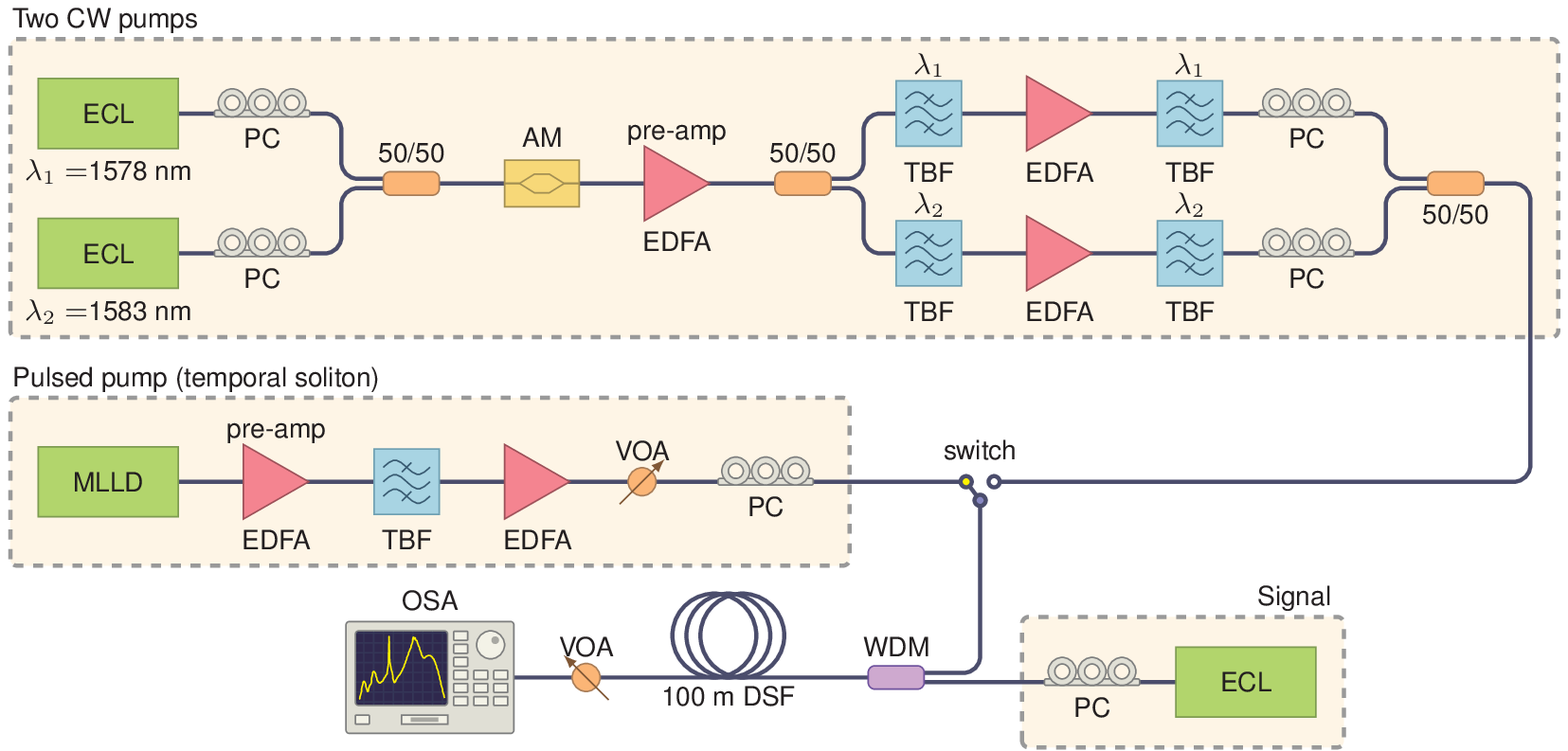}
\bcaption{Experimental set-up.}{ECL:~external cavity laser; PC:~polarization controller; AM:~amplitude modulator; EDFA:~erbium-doped fibre amplifier; TBF:~tunable bandpass filter; MLLD:~mode-locked laser diode; VOA:~variable optical attenuator; DSF:~dispersion-shifted fibre; OSA:~optical spectrum analyzer.}
\label{set-up}
\end{figure*}
\subsection{Four-wave mixing interpretation.} The above discussion highlights the equivalence of soliton interactions in nonlinear fibre optics\cite{yulin04,skryabin05, skryabin10} and fibre-optical analogues of event horizons\cite{philbin08}. However, neither of these descriptions provide information on the discrete nonlinear optical wave mixing mechanisms or energy conservation laws that drive the conversion of photons from the probe to the idler. We now address this gap by describing the frequency translation dynamics in terms of single-frequency electromagnetic fields. For this purpose, we consider two CW pumps at~$\omega_\pm$, with mean frequency~$\omega_\text{s}$ and separation~$\Delta$ such that $\omega_\pm=\omega_\text{s}\pm\Delta/2$, co-propagating with a weak probe at~$\omega_\text{p}$. Four-wave mixing between the three CW modes leads to the generation of a frequency comb with equidistant frequency components (Fig.~\ref{schematic}d). It has been shown\cite{xu13} that in the presence of higher-order dispersion, this cascade can be phase-matched such that a single higher-order idler mode at $\omega_\text{i}=\omega_\text{p}+n\Delta$ ($n=\pm1,\pm2,\dots$) is resonantly amplified. Here the positive and negative integers represent conversion from a low-frequency probe to a high-frequency idler and vice versa. We note that the intermediate frequency components (between $\omega_\text{p}$ and $\omega_\text{i}$) are not amplified, but instead simply serve to transfer energy to the resonant idler. In quantum-mechanical terms, the FWM cascade mimics a phase-matched higher-order nonlinear process in which $n$ pump photons at $\omega_\pm$ and a single probe photon $\omega_\text{p}$ are annihilated, while $n$ photons at $\omega_\mp$ and a single idler photon $\omega_\text{i}$ are created\cite{erkintalo12,xu13}, as shown in Fig.~\ref{schematic}c. Assuming the CW pumps $\omega_\pm$ to possess equal powers, the phase-matching condition leading to the amplification of the $n^\text{th}$ order idler is given by\cite{xu13}
\begin{equation}
\hat{D}(\omega_\text{i}-\omega_\text{s})=\hat{D}(\omega_\text{p}-\omega_\text{s})+n\left[\hat{D}(\Delta/2)-\hat{D}(-\Delta/2)\right].
\label{resonance2}
\end{equation}
Under typical conditions, the pump detuning $\Delta$ is small, so that \mbox{$\hat{D}(\Delta/2)\sim\hat{D}(-\Delta/2)$} (see top curve in Fig.~\ref{schematic}a).  In this case, the phase-matching condition is independent of the CW pump separation $\Delta$, and Eq.~\eqref{resonance2} reduces to the resonance condition describing event horizon dynamics, given in Eq.~\eqref{resonance1}. Therefore, all CW pairs $\omega_\pm$, regardless of their detuning $\Delta$, act to translate the frequency of the probe to the same idler frequency as illustrated in Fig.~\ref{schematic}d. A smaller pump detuning leads to an increased number of intermediate steps in the cascade, however the phase-matched idler frequency remains the same. The continuum limit converges to the soliton interaction described above\cite{erkintalo12}, and the formation of a fibre-optical event horizon now lends itself to an interpretation in terms of monochromatic CW fields: soliton modes, symmetrically distributed about the pump centre frequency, all drive the resonant amplification of the idler mode, which has a group velocity of opposite sign to the probe mode (in the reference frame of the soliton). Recalling that the FWM cascade also leads to an exchange of energy between the pump modes (see Fig.~\ref{schematic}c), this discrete description also reveals the energy conservation mechanism underlying the spectral recoil experienced by solitons during horizon reflections\cite{demircan11,demircan12,driben13,yulin13}, which is also visible in Fig.~\ref{schematic}b.

\begin{figure}[ht]
\centering
\includegraphics[clip=true]{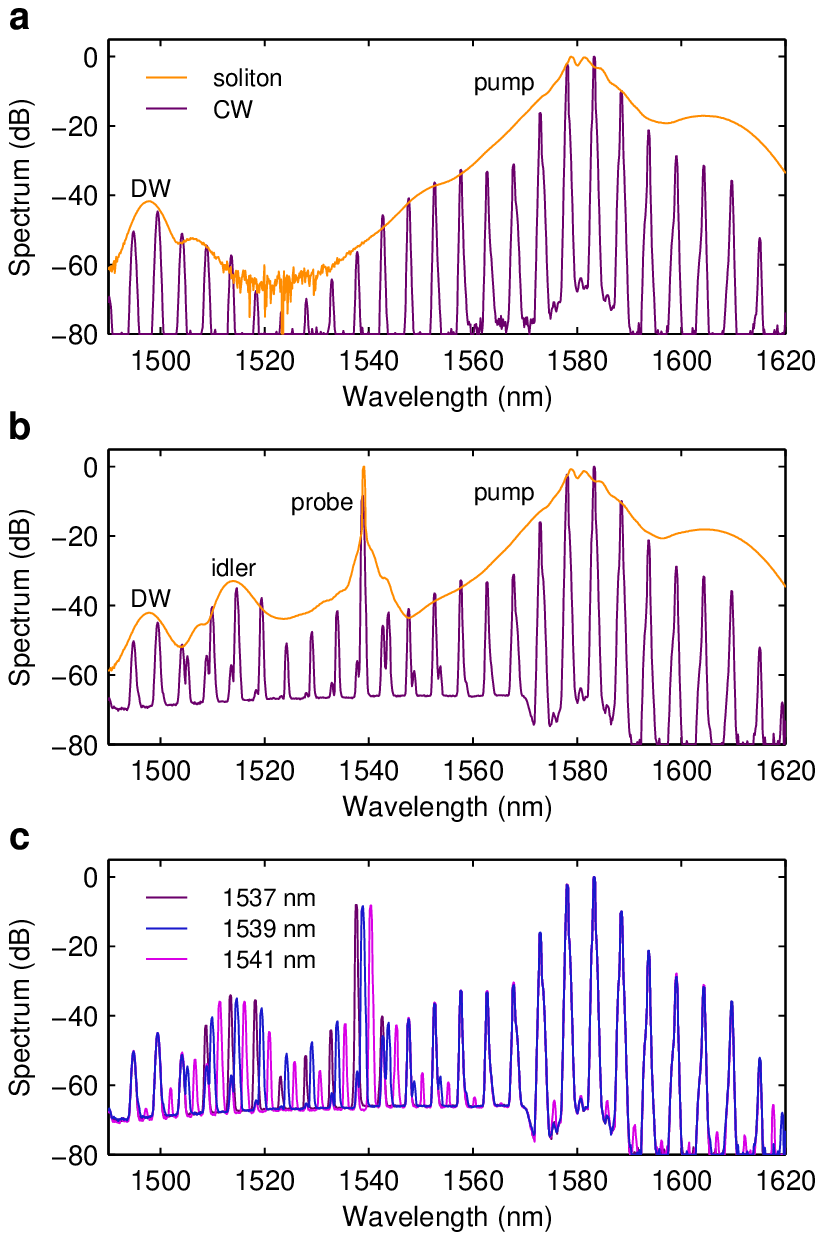}
\bcaption{First comparison of frequency translation induced by solitons and CW fields.}{(\textbf{a}) Output spectrum in the absence and (\textbf{b}) presence of a CW probe. Orange and purple curves correspond to soliton and CW pumps, respectively. (\textbf{c}) Output spectra for three different probe wavelengths (1537~nm, 1539~nm, and 1541~nm). Note that the 10~GHz frequency comb of the mode-locked soliton source is unresolved by the OSA.}
\label{exprfig1}
\end{figure}

\subsection{Experiments.} We now verify the above analysis by comparing the results of two sets of experiments. In the first experiment, we launch a temporal soliton into a segment of dispersion-shifted fibre (DSF) to create an artificial event horizon for a weak, frequency-tunable CW probe. In the second experiment we replace the soliton by two quasi-CW pumps, designed to mimic two spectral modes of the soliton. These experiments allow us to directly compare the frequency translation signatures of artificial event horizons, and the dynamics of cascaded four-wave mixing.

Figure~\ref{set-up} shows our experimental set-up (see also Methods). An amplified mode-locked laser diode serves as the source of picosecond solitons. The discrete CW pumps are derived from two frequency-tunable external cavity lasers (ECLs), which are amplitude modulated to produce flat-top nanosecond pulses.  Sufficient quasi-CW power is obtained by a two-stage amplification scheme, with narrow bandpass filters used to remove amplified spontaneous emission. Depending on the experimental configuration, the soliton or CW pumps are combined with a weak CW probe derived from a third ECL. The combined field is then launched into 100~m of DSF, and the spectrum is measured at the fibre output using an optical spectrum analyzer.

\begin{figure}[ht]
\centering
\includegraphics[clip=true]{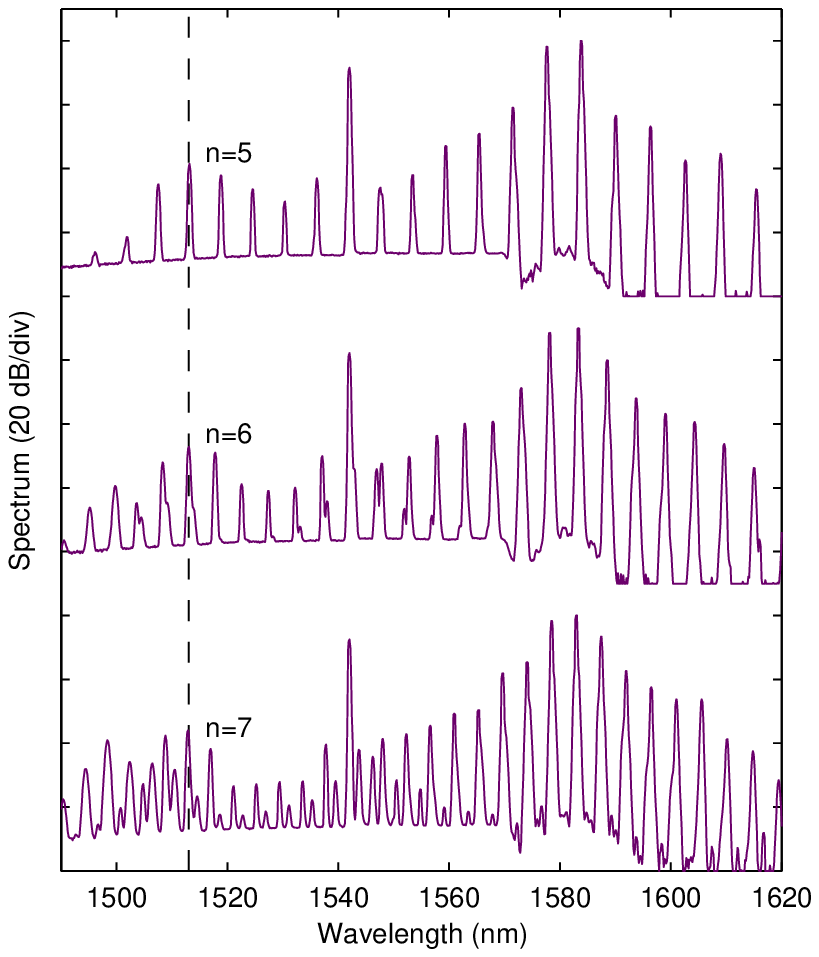}
\bcaption{Demonstration of the idler frequency insensitivity against CW pump detuning.}{Output spectra measured for CW pump detunings of (from top to bottom) $\Delta/(2\pi)\sim0.74$~THz, $\Delta/(2\pi)\sim0.62$~THz, and $\Delta/(2\pi)\sim0.53$~THz. The corresponding cascade orders are $n=5$, $n=6$, and $n=7$. Dashed black line indicates the wavelength phase-matched through the ``event horizon'' resonance.}
\label{exprfig1.5}
\end{figure}
\subsection{Comparison of results.} Figure~\ref{exprfig1}a superimposes output spectra for both pump configurations (dual-CW and soliton pulse) in the absence of the probe wave. The soliton emits a dispersive wave centred at 1500~nm, which is closely matched by the cascaded FWM components generated by the CW pumps \cite{erkintalo12}. When the weak probe at 1539~nm is also injected into the fibre, an idler component, blue-shifted relative to the probe, appears at 1515~nm, shown in Fig.~\ref{exprfig1}b. This is the frequency-domain signature of a fibre-optic white-hole horizon \cite{philbin08}. In the time domain, the probe is travelling faster than the soliton, leading to the generation of an idler which travels slower than the soliton (see Fig.~\ref{schematic}a). The generation of the idler is observed for both pumping configurations, with the continuous soliton spectrum closely following the frequency comb generated by the CW pumps. Indeed, the agreement between the two experiments is remarkable.

We emphasize that the processes generating the DW and the idler are completely independent, which is highlighted in Fig.~\ref{exprfig1}c. Here we show three different output spectra for varying probe wavelengths. The amplified idler mode changes as the probe is varied, in agreement with Eq.~\eqref{resonance2}, but the frequency components corresponding to the DW remain unaffected. In Fig.~\ref{exprfig1.5}, we experimentally demonstrate the insensitivity of the phase-matched idler frequency to the pump detuning $\Delta$ (shown schematically in Fig.~\ref{schematic}d). Here, we have selected comparatively large detunings for which the bracketed term in Eq.~\eqref{resonance2} may not completely vanish. Despite choosing low cascade orders $n$ ($n\propto1/\Delta$), the wavelengths of the phase-matched idler components differ by less than a nanometre, and agree well with the computed ``event horizon'' limit from Eq.~\eqref{resonance1}. Our experiments thus confirm the predicted frequency invariance of the amplified idler, and agree well with the trend expected for the limit $\Delta\rightarrow0$.

\begin{figure*}[t!]
\centering
\includegraphics[clip=true]{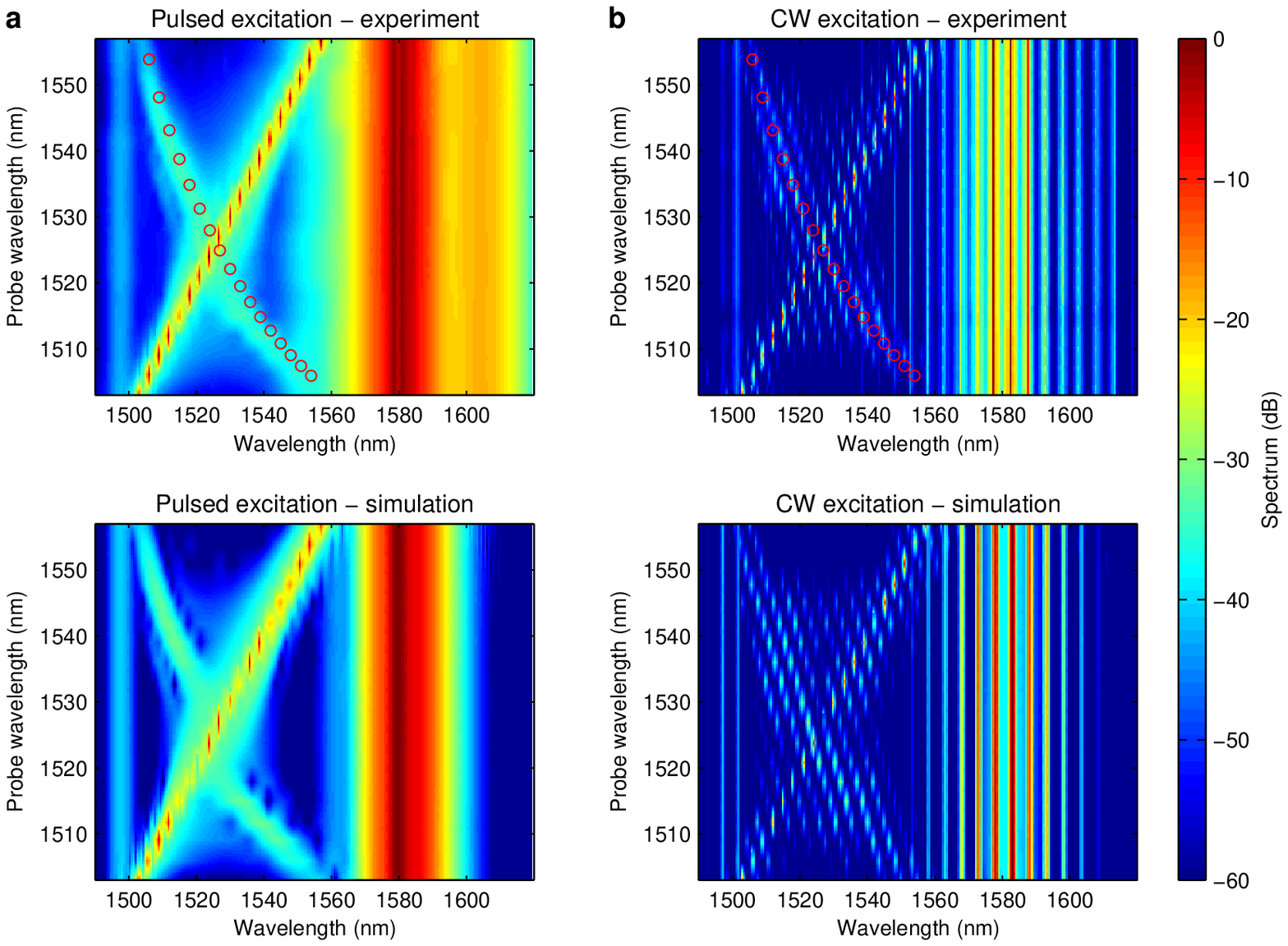}
\bcaption{Comparison of dynamics induced by solitons and CW fields over the whole range of observable frequency conversion.}{(\textbf{a, b})  Experimental (top) and numerically simulated (bottom) output spectra for (\textbf{a}) pulsed and (\textbf{b}) CW excitation as the probe wavelength is varied. Open circles in the top graphs indicate idler wavelengths predicted by theory.}
\label{exprfig2}
\end{figure*}
\subsection{Full event horizon dynamics.} In order to unambiguously establish the validity of our FWM description, we have used both pumping configurations to explore the dynamics over the full range of probe frequencies that display event horizon signatures. The density maps in the top row of Fig.~\ref{exprfig2} show a sequence of experimental spectra corresponding to pulsed and CW pumping. In both pumping regimes, the probe is varied between the boundary limits set by the pump and the DW wavelengths. We first note that the DW component remains entirely unaffected as the probe wavelength is scanned, further confirming its independence from the event horizon process. Secondly, in both experiments the probe-idler pairs paint an apparent \emph{X}-shaped feature as the probe frequency is varied, owing to the interchangeability of the probe and the idler. The two intersect at 1526~nm, which corresponds to the wavelength that is group-velocity matched with the pump. When the probe is tuned through this point, the dynamics are reversed, since the group velocity matched point represents the transition point between a white- and black-hole horizon. The agreement between the temporal soliton and discrete CW experiments is remarkable, which acts as a clear demonstration of their correspondence. For each probe wavelength, we have also calculated the idler wavelength as predicted theoretically by the resonance condition given by Eq.~\eqref{resonance1}, and observe near perfect agreement. Finally, we have also performed full numerical modelling of these experiments (see Methods), shown in the bottom row of Fig.~\ref{exprfig2}, and the simulation results can be seen to be in excellent agreement with the experiments. Small discrepancies are attributed to imperfect knowledge of fibre parameters, and to residual amplified spontaneous emission seen on the red-edge of the soliton experiments. Numerical simulations further confirm that no frequency translation takes place for probe frequencies larger than the DW limit, in line with the phase-matching conditions (see also Fig.~\ref{schematic}a).\\


\section*{Discussion}
In this paper, we have described the nonlinear optics of fibre event horizons. First, we have clarified how frequency conversion mechanisms extensively studied in the context of ultrafast nonlinear fibre optics describe the very same physics as fibre event horizons. In particular, the nonlinear interaction of a weak probe with an intense soliton results in generation of a frequency-shifted idler, whose group velocity relative to the soliton is opposite to that of the probe's. We have then demonstrated how this mechanism can be completely described within the discrete wave mixing framework of traditional nonlinear optics. Specifically, the cascaded four-wave mixing of two strong CW modes and a weak probe results in the resonant amplification of a higher-order idler, whose group velocity relative to the soliton is opposite to that of the probe's. In the continuum limit, where the frequency spacing of the CW pumps approaches zero, the phase-matching condition governing the FWM interactions reduces to the resonance condition governing the soliton dynamics, illustrating the correspondence of the two phenomena. Extensive experiments designed to compare spectral signatures of horizon dynamics at all possible probe wavelengths, induced by both a temporal soliton and by a pair of quasi-CW pumps, confirm our analysis.

Our work establishes a formal link between artificial event horizons, ultrafast nonlinear fibre-optics, and four-wave mixing of monochromatic continuous waves, and we believe that our results will permit valuable flow of knowledge, ideas and insights across all of these fields. As an example, it is known that higher-order dispersion is critical for the generation of new frequencies through the soliton and CW processes discussed above\cite{yulin04,skryabin05, skryabin10, xu13}, implying the same requirement for the observation of fibre-optic event horizons. Finally, because our FWM description allows for the photon creation and annihilation processes to be identified, our results could represent a significant step towards the unambiguous observation of fibre-optic photon pair creation analogous to the elusive Hawking radiation\cite{belgiorno10,barcelo11, petev13, unruh14}.\\


\section*{Methods}
{\small{\subsection{Experimental set-up.} The CW pumps are derived from two tunable fibre-coupled external-cavity lasers (ECLs) with wavelengths set in the telecommunications L-band (centre wavelength $\lambda_\text{s}=1580.46$~nm). They are combined using a 50:50 fibre coupler, before being passed through an amplitude modulator to produce 1~ns flat-top pulses with a 100:1 duty cycle. The output is pre-amplified by a low-noise L-band erbium-doped fibre amplifier (EDFA), before being split with a second 50:50 fibre coupler, and amplified by two separate high-power L-band EDFAs in order to avoid unwanted wave mixing between the two pumps. The pumps are spectrally filtered before and after the high-power amplification stage by 0.4~nm tunable bandpass filters in order to remove unwanted amplified spontaneous emission. The two pumps are recombined with a final 50:50 fibre coupler, resulting in two 1~ns flat-top, quasi-CW pulses with a peak power of 5~W each. The extra port of the last coupler is used to monitor the pump power, and ensure the pulses are temporally synchronized. The CW pumps and a 20~mW continuous-wave signal, derived from a third ECL, are combined using a wavelength-division multiplexer (WDM). Polarization controllers before the WDM ensure both the pumps and signal are co-polarized. The combined field is then launched into 100~m of dispersion-shifted fibre (Corning DSF). The output of the fibre is attenuated using a variable optical attenuator before being recorded by an optical spectrum analyzer (Ando AQ-6315A). For the temporal soliton experiment, the dual-CW pump is replaced with a 10~GHz mode-locked laser diode (Alnair MLLD-100) producing 1~ps pulses, set to the same centre wavelength as the CW pumps. The pulses are again amplified in two stages, with a single filter between the two amplifiers. An attenuator before the WDM allows us to control the peak power of the solitons. The peak power of the pulses launched into the fibre is 17.2~W.

\subsection{Numerical simulations.} Our experimental observations are modelled by simulations based on the generalized nonlinear Schr{\"{o}}dinger equation (NLSE)\cite{dudley06}
\begin{equation}
\frac{\partial A}{\partial z}=\sum_{k\geq2}\frac{i^{k+1}}{k!}\beta_k\frac{\partial^kA}{\partial t^k}+i\gamma A(z,t)\!\!\!\int\limits_{-\infty}\limits^{+\infty}\!\!\!R(t')|A(z,t-t')|^2dt',
\label{nlse}
\end{equation}
where $A(z,t)$ is the envelope of the pulse, $\beta_k$ are the dispersion coefficients, $\gamma$ is the nonlinear interaction coefficient of the fibre, and $R(t)=(1-f_R)\delta(t)+f_Rh_R(T)$ is the response function containing the instantaneous and delayed Raman contributions to the nonlinearity. The Raman response function $h_R(t)$ is experimentally measured\cite{stolen89}, and $f_R=0.18$ is the fractional Raman contribution to the nonlinearity. The NLSE is solved numerically using the split-step Fourier method.\\
\indent In Fig.~\ref{schematic}(b), we show an illustrative simulation of the temporal and spectral evolution of a soliton pump and a weak probe pulse propagating in 50~m of optical fibre. The parameters used in this illustrative simulation are chosen to clearly show the frequency translation $\omega_\text{p}\rightarrow\omega_\text{i}$, and do not match experimental parameters. The simulation uses dispersion coefficients $\beta_2=-4.4$~ps$^2$km$^{-1}$ and $\beta_3=+0.13$~ps$^3$km$^{-1}$ at the pump, and an estimated nonlinear interaction coefficient $\gamma=2.5$ W$^{-1}$km$^{-1}$. Raman scattering was neglected in this simulation. The soliton pump is centred at 1600~nm, with peak power $P_\text{s}=245$~W and pulse width $\tau_\text{s}=85$~fs, while the probe is centred at 1522~nm and has peak power $P_\text{p}=5$~W and pulse width $\tau_\text{p}=300$~fs. The probe trails the pump temporally by 1.1~ps at the input of the fibre.\\
\indent For the simulations shown in Fig.~\ref{exprfig2}c-d, we use fibre dispersion coefficients $\beta_2=-2.6$~ps$^2$km$^{-1}$, $\beta_3=+0.13$~ps$^3$km$^{-1}$, and $\beta_4=-7\times10^{-4}$~ps$^4$km$^{-1}$ at the pump centre wavelength, and nonlinear coefficient $\gamma=2.5$ W$^{-1}$km$^{-1}$. These simulations also included stimulated Raman scattering for completeness, but its effect was found to play a negligible role in the dynamics in our parameter regime. For all simulations, we have also checked that the impact of self-steepening is negligible. The optical power and wavelengths of the pump and signal waves are the same as those measured in experiments. The temporal width of the soliton pump is $\tau_\text{s}=1.5$~ps, estimated from a second-harmonic frequency-resolved optical gating (Southern Photonics SH-150) measurement.}}\\



\vspace*{20pt}

{\small{
\section*{Acknowledgements}
M. E., N. G. B. and S. G. M. acknowledge support from the Marsden Fund of the Royal Society of New Zealand. J. M. D. acknowledges support from the European Research Council (ERC) Advanced Grant ERC-2011-AdG-290562 MULTIWAVE. G. G. acknowledges support from the Academy of Finland (projects 130099 and 132279).\\

\section*{Author Contributions}
K. E. W. performed all experiments and numerical simulations; Y. Q. X. developed the initial experimental set-up; M. E. and G. G. developed the concept and the theory of the frequency domain description of horizon dynamics. N. G. B. and J. M. D. contributed to clarifying the soliton-horizon dynamics concepts; S. G. M. designed the experiment, supervised the work and obtained financial support. All authors contributed to analyzing results, writing and commenting on the manuscript and discussing the results and interpretation.\\

\section*{Competing Financial Interests}
The authors declare no competing financial interests.
}}

\end{document}